\documentclass[hidelinks,conference]{IEEEtran}
\usepackage{amsmath,amsfonts}
\usepackage{algorithmic}
\usepackage{algorithm}
\usepackage{array}
\usepackage[caption=false,font=normalsize,labelfont=sf,textfont=sf]{subfig}
\usepackage{textcomp}
\usepackage{stfloats}
\usepackage{url}
\usepackage{verbatim}
\usepackage{graphicx}
\usepackage{cite}
\usepackage{amssymb}
\usepackage{bm}
\usepackage{caption}
\usepackage{multirow}
\usepackage{makecell}
\usepackage{tabularx}
\usepackage{xcolor}
\usepackage{orcidlink}
\usepackage[acronym,shortcuts]{glossaries}

\newacronym{AMP}{AMP}{approximate message passing}
\newacronym{GaBP}{GaBP}{Gaussian belief propagation}
\newacronym{SotA}{SotA}{state-of-the-art}
\newacronym{SE}{SE}{state evolution}
\newacronym{i.i.d.}{i.i.d.}{independent and identically distributed}
\newacronym{ML}{ML}{maximum likelihood}
\newacronym{BNN}{BNN}{binary neural network}
\newacronym{RAMP}{RAMP}{regularized approximate message passing}
\newacronym{RGaBP}{RGaBP}{regularized Gaussian belief propagation}
\newacronym{IDLS}{IDLS}{iterative discrete least squares}
\newacronym{ZF}{ZF}{zero forcing}
\newacronym{LMMSE}{LMMSE}{linear minimum mean squared error}
\newacronym{AWGN}{AWGN}{additive white Gaussian noise}
\newacronym{QT}{QT}{quadratic transform}
\newacronym{MAP}{MAP}{maximum a posteriori}
\newacronym{QAM}{QAM}{quadrature amplitude modulation}
\newacronym{QPSK}{QPSK}{quadrature phase shift keying}
\newacronym{PAM}{PAM}{pulse amplitude modulation}
\newacronym{DA-ZF}{DA-ZF}{discrete-aware zero forcing}
\newacronym{DA-LMMSE}{DA-LMMSE}{discrete-aware linear minimum mean squared error}
\newacronym{CS}{CS}{compressed sensing}
\newacronym{OAMP}{OAMP}{orthogonal approximate message passing}
\newacronym{VAMP}{VAMP}{vector approximate message passing}
\newacronym{SK}{SK}{Sherrington–Kirkpatrick}
\newacronym{wlg}{wlg}{without loss of generality}
\newacronym{MIMO}{MIMO}{multiple-input multiple-output}
\newacronym{MSE}{MSE}{mean squared error}
\newacronym{BER}{BER}{bit error rate}

\begin{document}

\title{Regularized Approximate Message Passing for Overloaded Discrete Linear Inversion
\vspace{-0.5ex}}

% \author{Shreesal Shrestha, Getuar Rexhepi, Kuranage Roche Rayan Ranasinghe, Hyeon Seok Rou \\ and Giuseppe Thadeu Freitas de Abreu
% % <-this % stops a space 
% \thanks{S. Shrestha, K. R. R. Ranasinghe, H. S. Rou, and G. T. F. de Abreu are with the School of Computer Science and Engineering, Constructor University (previously Jacobs University Bremen), Campus Ring 1, 28759 Bremen, Germany (emails: \{shrshresth,kranasinghe, hrou, gabreu\}@constructor.university).}}

\author{
\IEEEauthorblockN{Shreesal Shrestha$^*$, Getuar Rexhepi$^*$, Kuranage Roche Rayan Ranasinghe$^*$, \\ Hyeon Seok Rou$^*$ and Giuseppe Thadeu Freitas de Abreu$^*$}
\IEEEauthorblockA{$^*$\textit{School of Computer Science and Engineering, Constructor University, 28759 Bremen, Germany} \\
Emails: [shrshresth,grexhepi,kranasinghe, hrou, gabreu]@constructor.university}
\vspace{-5ex}
}


\maketitle

\begin{abstract}
We propose \ac{RAMP}, a low-complexity algorithm for discrete signal detection in overloaded \ac{MIMO} systems where the number of transmit antennas exceeds the number of receive antennas. 
While the \ac{SotA} \ac{IDLS} framework achieves near-optimal discrete-aware performance, its iterative matrix inversions impose a prohibitive $\mathcal{O}(M^3)$ complexity. 
\ac{RAMP} resolves this by deriving an adaptive, state-dependent scalar denoiser that enforces arbitrary discrete constellation constraints within the \ac{AMP} framework, reducing per-iteration complexity to $\mathcal{O}(NM)$. 
A robust variant is further proposed by incorporating an $\ell_2$-norm penalty, analogous to a \ac{LMMSE} estimator, to enhance noise resilience.
Simulation results under uncorrelated Rayleigh fading demonstrate that both proposed algorithms closely track their exact \ac{IDLS} counterparts while avoiding the catastrophic failure of standard \ac{AMP} in the overloaded regime, achieving steep \ac{BER} waterfall curves at a fraction of the computational cost.

\end{abstract}

\begin{IEEEkeywords}
AMP, IDLS, overloaded, message passing.
\end{IEEEkeywords}

\glsresetall

% \vspace{-1ex}
\section{Introduction}
\IEEEPARstart{T}{he} reconstruction of high-dimensional discrete signals from noisy, underdetermined linear measurements arises in numerous scientific and engineering applications.
This problem is particularly relevant in the realm of wireless communications, where the adoption of \ac{MIMO} technologies has enabled the transmission of multiple data streams over the same time-frequency resources, significantly improving spectral efficiency and reliability \cite{yang2015mimo}. 

The central difficulty arises in the overloaded/under-determined regime where the number of unknowns $M$ exceeds the number of observations $N$.
In classical estimation theory, such a system is called ill-posed, as the channel matrix possesses a non-trivial null space leading to infinitely many solutions that satisfy the linear constraints.
Consequently, standard linear reconstruction techniques, such as \ac{ZF} and \ac{LMMSE}, are rendered ineffective. 
Such methods recover a projection of the signal that is contaminated by null-space components, which results in an irreducible error floor.

To resolve this, one can exploit the discreteness of the constellations of the signal.
However, while the discrete constraint makes the problem theoretically unique, the search space grows exponentially with $M$, rendering the optimal \ac{ML} detector NP-hard.
Unlike \ac{CS} \cite{FoucartRauhut2013, Das2013}, which relies on the signal being sparse, the problem considered here involves signals that are dense but constrained to a finite, discrete constellation $\mathcal{S}$ (i.e., $\bm{s} \in \mathcal{S}^M$).
When the signal is dense, standard \ac{CS} relaxation techniques such as $\ell_1$-minimization provide a loose approximation for the discrete $\ell_0$ constraint, resulting in a significant performance degradation \cite{Hayakawa2017, Hayakawa2018}.
Additionally, low-complexity solutions such as \ac{AMP} \cite{Donoho2009, Bayati2011, Montanari2011}, originally developed for \ac{CS}, tend to fail in dense underdetermined systems due to the ill-posed nature of the problem.

The \ac{IDLS} framework \cite{iimori2020robust} was recently proposed as a high-performance solution to this problem, albeit at a steep computational cost. 
The iterative solution of the proposed framework requires a matrix inversion at every step, leading to a cubic complexity of $\mathcal{O}(M^{3})$.
To address this computational bottleneck, we introduce a message-passing algorithm with an adaptive, state-dependent scalar denoiser that enforces arbitrary discrete constellation constraints based upon an \ac{AMP} framework termed \ac{RAMP}.
By utilizing both the \ac{IDLS} framework from \cite{iimori2020robust} and principles of \ac{AMP}, \ac{RAMP} lowers the per-iteration complexity from cubic $\mathcal{O}(M^{3})$ to quadratic $\mathcal{O}(MN)$.

\section{System Model and Problem Formulation}
We consider a linear vector system model, characteristic of an overloaded \ac{MIMO} uplink system, where the received vector $\bm{y} \in \mathbb{C}^{N}$ is modeled as
\begin{equation}
    \bm{y}=\bm{H}\bm{s}+\bm{n},
    \label{eqn:system_model}
\end{equation}
where $\bm{H} \in \mathbb{C}^{N \times M}$ is the channel matrix, with $N$ receive antennas and $M$ transmit antennas. 

The system is overloaded/under-determined, meaning $M > N$. The noise $\bm{n} \in \mathbb{C}^{N}$ is an \ac{i.i.d.} circularly symmetric complex \ac{AWGN} vector, with $\bm{n} \sim \mathbb{CN}(0, \sigma^2_n \bold{I}_{N})$. 
The transmit vector $\bm{s} \in \mathbb{C}^{M}$ is the vector of transmitted symbols, which are drawn from a discrete modulation alphabet $\mathcal{S}$ of either \ac{QAM} of size $K = 2^b$, where $b$ is the number of bits per symbol, for complex systems or of \ac{PAM} of size $K=2^{b/2}$ for real systems.

Given \eqref{eqn:system_model}, the \ac{ML} detection problem of the complex transmit signal vector $\bm{s} \in \mathcal{S}^{M}$ can be expressed as an $\ell_2$ norm minimization problem
\begin{equation}
    \hat{\bm{s}}_{\mathrm{ML}} = \arg \min_{\bm{s} \in \mathcal{S}^{M}} \|\bm{y} -\bm{H}\bm{s}\|^2_2.
    \label{eqn:ML_solution}
\end{equation}

To avoid the NP-hard search of \ac{ML} detection, the \ac{IDLS} framework in \cite{iimori2020robust} reformulates this discrete optimization problem into a tractable, continuous one.
The method starts with replacing the hard constraint $\bm{s} \in \mathcal{S}^{M}$ with a penalty term based on a continuous, smooth approximation of the $\ell_0$-norm \cite[eq. 10]{iimori2020robust}, resulting in a non-convex objective function. This can then be convexified via fractional programming \cite{Shen2018}, whereby the adaptive per-symbol weights $\beta_{i,m}$ are computed at each iteration $t$ as
\begin{equation}
    \beta_{i,m}^{(t)} = \frac{\sqrt{\alpha}}{|\hat{s}_m^{(t-1)} - c_i|^2 + \alpha},
    \label{eqn:beta}
\end{equation}
where $\hat{s}_m^{(t-1)}$ is the estimate of the $m$-th symbol from the previous iteration, $c_i \in \mathcal{S}$ is a constellation point, and $\alpha$ is a small constant controlling the approximation's tightness.

This procedure transforms the original non-convex problem into a simple, convex quadratic minimization.
The ``base case'' \ac{IDLS} formulation, which we solve here under a \ac{RAMP} framework in Section \ref{section:RAMP}, is given by \cite[eq. 16]{iimori2020robust}
\begin{equation}
    \arg \min_{\bm{s} \in \mathbb{C}^{M}} \|\bm{y} - \bm{Hs}\|^2_2 + \lambda\sum^{2b}_{i=1}\sum^{M}_{m=1} \beta_{i,m}^2 |s_m-c_i|^2,
    \label{eqn:IDLS_Base_Case}
\end{equation}
where $\lambda$ is a regularization parameter which can be optimally set following \cite[eq. 41]{iimori2020robust} or other methods such as those in \cite{Boukari1995, Pedregosa2016, Ehrhardt2024, Bischl2023}.

The \ac{IDLS} framework  also offers a ``robust'' formulation analogous to the \ac{LMMSE} estimator, by adding an $\ell_2$-norm penalty on the signal $\bm{s}$ itself to improve the noise resilience \cite[eq. 48]{iimori2020robust}
\begin{equation}
    \arg \min_{\bm{s} \in \mathbb{C}^{M}} \|\bm{y} - \bm{Hs}\|^2_2 + \sigma^2_n \| \bm{s}\|_2^2 + \lambda\sum^{2b}_{i=1}\sum^{M}_{m=1} \beta_{i,m}^2 |s_m-c_i|^2.
    \label{eqn:IDLS_Robust_Case}
\end{equation}

Note that this addition specifically provides robustness against noise amplification (similar to the \ac{LMMSE}) and does not inherently address the structural ill-conditioning caused by spatial correlation.

Defining a vector $\bm{b}$ and the diagonal matrix $\bm{B}$ constructed from the weights $\beta_{i,m}$, the closed form solution to the ``robust'' problem \eqref{eqn:IDLS_Robust_Case} is given by \cite[eq. 50a]{iimori2020robust}
\begin{equation}
    \hat{\bm{s}} = (\bm{H}^H + \sigma_n^2\bm{I}_{N} + \lambda\bm{B})^{-1}(\bm{H}^H\bm{y}+\lambda\bm{b}).
    \label{eqn:robust_idls_solution}
\end{equation}

However, despite the excellent discrete-aware performance offered by \eqref{eqn:robust_idls_solution}, the per-iteration direct matrix inversion establishes a prohibitive $\mathcal{O}(M^{3})$ complexity.
In the following sections, we propose a \ac{RAMP} framework to solve the same optimization problems while completely avoiding this matrix inversion and reducing the complexity to an efficient $\mathcal{O}(NM)$.

\section{Regularized Approximate Message Passing}
\label{section:RAMP}

We seek to replace the high complexity inversion \eqref{eqn:robust_idls_solution} by minimizing the complexity, yet still achieving similar performance. 
To this end, in this section we propose a \ac{RAMP} solution as a low complexity solver for overloaded systems.

\subsection{RAMP Framework}
To develop a low-complexity solver, we first seek to reformulate the deterministic \ac{IDLS} problem \eqref{eqn:IDLS_Base_Case} as a probabilistically consistent \ac{MAP} estimation problem.
Given the \ac{AWGN} channel $\bm{y} = \bm{H}\bm{s} + \bm{n}$ with $\bm{n} \sim \mathcal{CN}(0, \sigma^2_n)$, we scale the total negative log-probability by $1/\sigma^2_n$ such that
\begin{equation}
    \min_{\bm{s} \in \mathbb{C}^{M}} 
    \underbrace{ \frac{1}{\sigma_n^2} \|\bm{y} - \bm{H}\bm{s}\|_2^2 }_{\propto -\log \mathrm{P}(\bm{y}|\bm{s})} + 
    \underbrace{ \lambda_\mathrm{eff} \sum_{i=1}^{2^b}\sum_{m=1}^{M} \beta_{i,m}^2 |s_m - c_i|^2 }_{\propto -\log \mathrm{P}(\bm{s})},
    \label{eqn:normalized_cost_func}
\end{equation}
where $\lambda_\mathrm{eff} \triangleq \lambda/\sigma^2_n$. 

The scaling by the noise variance $\sigma^2_n$ frames the regularizer $\lambda_\mathrm{eff}$ as a strength relative to the noise power and provides the correct formulation for the derivation of the proposed denoiser in Section \ref{subsection:RAMP Denoiser}. 

Our proposed \ac{RAMP} algorithm builds upon the well-established \ac{AMP} framework \cite{Donoho2009, Bayati2011,Montanari2011}. 
\ac{AMP} is a \ac{SotA} iterative solver for large-scale linear inverse problems which decouples the vector matrix equation into $M$ parallel scalar problems through an iterative process of effective observation and denoising.

At each iteration $t$, the standard \ac{AMP} algorithm proceeds in the following steps:

\textbf{Effective Observation} : An effective observation $\bm{r}^{(t)} \in \mathbb{C}^{M}$ is formed by adding the current estimate $\hat{\bm{s}}^{(t)}$ to the matched-filtered residual $\bm{z}^{(t)}$
\begin{equation}
    \bm{r}^{(t)} = \hat{\bm{s}}^{(t)} + \bm{H}^H\bm{z}^{(t)}.
    \label{eqn:effective_obs}
\end{equation}

An important property of \ac{AMP} to note is the Onsager correction term, defined in \eqref{eqn:residual_upd}. 
As proven for standard \ac{AMP} (i.e., with \ac{i.i.d.} Gaussian matrices and static denoisers) \cite{Donoho2009,Bayati2011,Montanari2011}, the Onsager term perfectly decorrelates the current effective observation from the previous iterates in the large system limit. 
This perfect decorrelation is precisely what leads the components of $r_j^{(t)}$ to become statistically equivalent to the true symbol $s_j$ corrupted by \ac{i.i.d.} \ac{AWGN} $\mathcal{CN}(0, \sigma_t^2)$.

The variance of this effective noise $\sigma_t^2$ can be precisely tracked by a scalar recursion called \ac{SE}
\begin{align}
    \sigma^2_{t+1} &= \sigma_n^2 + \frac{1}{\delta}\text{MSE}(\eta_t,\sigma_t^2),
    \label{eqn:SE}
\end{align}
where $\delta = {N}/{M}$ is the system aspect ratio. 

\vspace{-2ex}
\begin{figure}[H]
    \centering
    \includegraphics[width=\columnwidth]{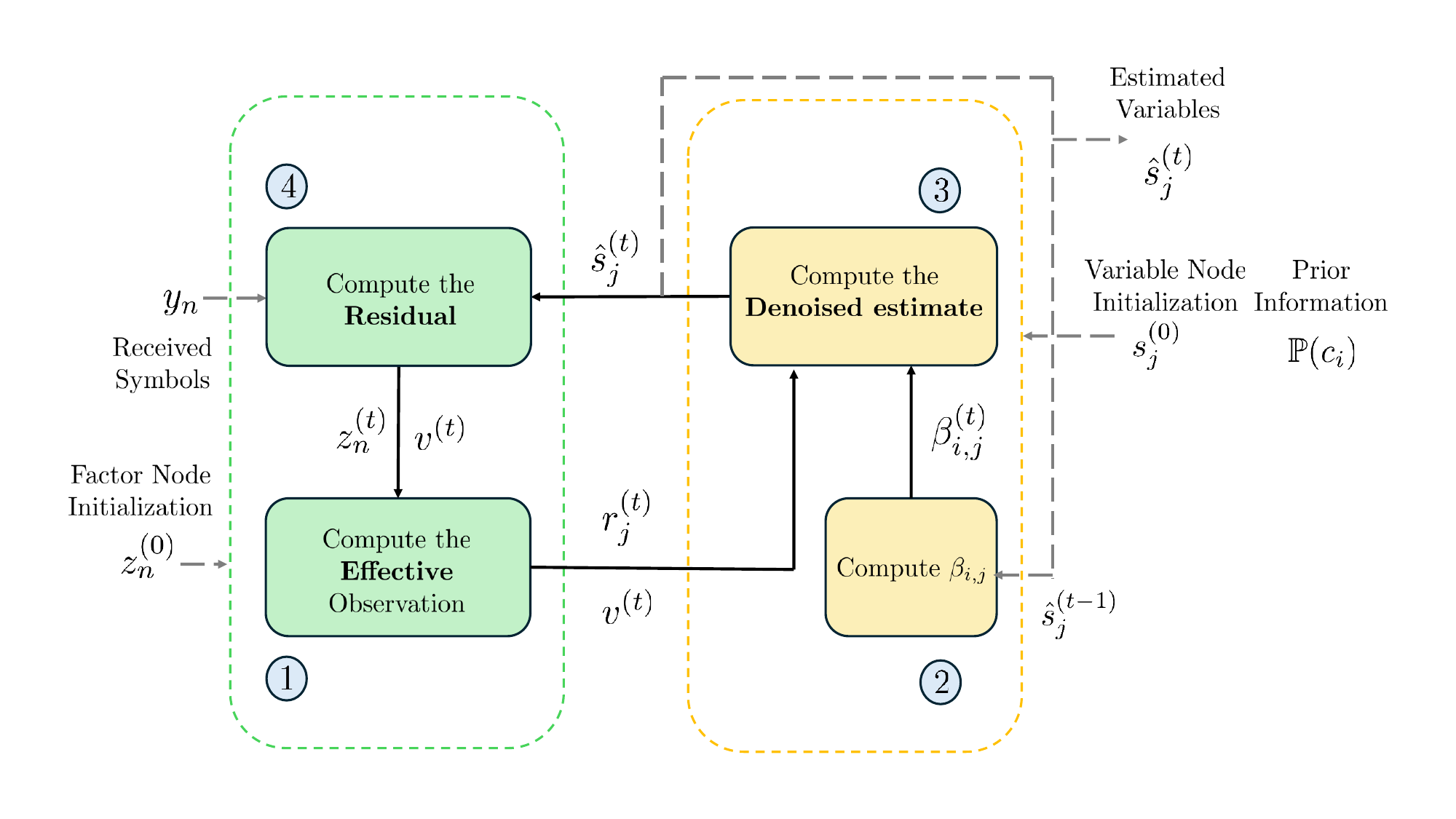}
    \vspace{-5ex}
    \caption{Schematic of the proposed \ac{RAMP} detector.}
    \label{fig:flowchart_amp}
\end{figure}

\begin{figure}[H]
    \centering
    \includegraphics[width=\columnwidth]{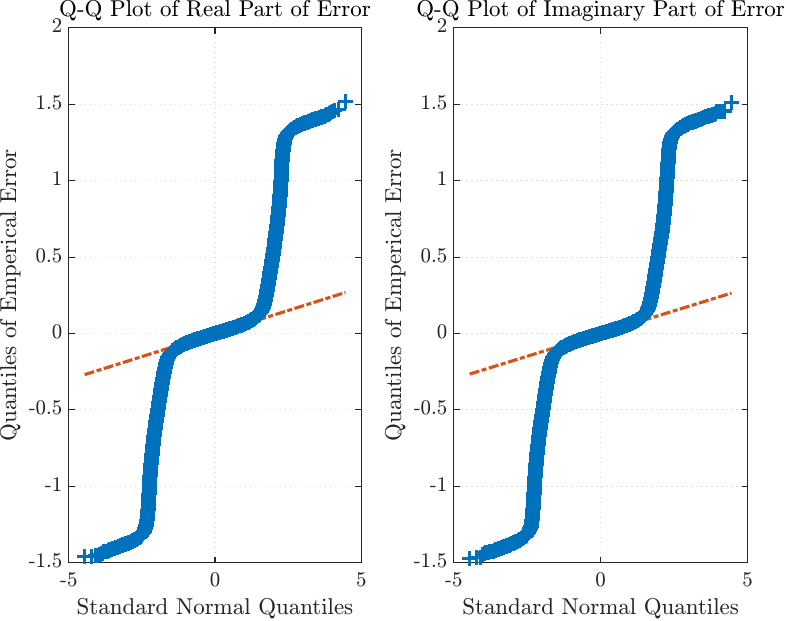}
    \caption{QQ-Plot of the internal error distribution ($M=120$, $N=96$, $E_b/N_0 = 10$ dB, QPSK)}
    \label{fig:QQ_plot}
\end{figure}
    
Although the \ac{SE} perfectly tracks the variance of the effective noise $\sigma_t^2$, we can also use the power of the residual as its consistent estimator \cite{Bayati2011,Montanari2011}
\begin{equation} 
    v^{(t)}  = \frac{1}{N}\|\bm{z}^{(t)}||_2^2.
    \label{eqn:variance_upd}
\end{equation}

\textbf{Scalar Denoising} : Given the effective observation $\bm{r}^{(t)}$, the large-matrix estimation problem is reduced to a set of parallel scalar denoising problems.
A non-linear denoising function $\eta(.,.)$ is applied element-wise to $\bm{r}^{(t)}$ to produce the new estimate $\hat{\bm{s}}^{(t+1)}$
\begin{equation}
    \hat{\bm{s}}^{(t+1)} = \eta(\bm{r}^{(t)}, v^{(t)}).
\end{equation}

\textbf{Residual Update} : The residual $\bm{z}^{(t+1)} \in \mathbb{C}^{N}$ is updated, incorporating the previously mentioned \textbf{Onsager correction term}, which enables the decoupling described above
\begin{equation}
    \bm{z}^{(t+1)} = \bm{y} - \bm{H}\bm{\hat{s}}^{(t+1)} + \underbrace{\frac{1}{\delta}\bm{z}^{(t)}\langle \eta'(\bm{r}^{(t)}, v^{(t)}) \rangle}_{\text{Onsager Term}},
    \label{eqn:residual_upd}
\end{equation}
where $\langle \eta'(\cdot) \rangle = \frac{1}{M} \sum_{m=1}^{M} \frac{\partial \eta_m}{\partial r_m}$ is the average divergence of the denoiser.

\subsection{RAMP Denoiser}
\label{subsection:RAMP Denoiser}
In standard applications like \ac{CS}, $\eta(\cdot)$ is a fixed denoising.
However, for the discrete overloaded problem, we require a denoiser that adaptively enforces constellation constraints.
Thus, we derive $\eta(\cdot)$ as the minimizer of the scalar version of our probabilistically consistent normalized \ac{IDLS} function \eqref{eqn:normalized_cost_func}
\begin{equation}
    \arg \min_{s_m} \frac{1}{v^{(t)}}|r_m^{(t)} - s_m|^2 + \lambda_\mathrm{eff} \sum_{i=1}^{2^b} (\beta_{i,m}^{(t)})^{2} |s_m - c_i|^2.
\end{equation}

The Wirtinger derivative w.r.t. $s_m^*$ can then be expressed as
\begin{equation}
    {\frac{\partial}{\partial{\bm{s_m^*}}} = \frac{1}{v^{(t)}}(s_m - r_m) + \lambda_\mathrm{eff}\sum_{i=1}^{2^b}(\beta_{i,m}^{(t)})^2(s_m-c_i)}.
\end{equation}

Solving for \(s_m\) yields
\begin{equation}
    \hat{s}_m^{(t+1)} = \frac{r_m^{(t)}+\lambda_\mathrm{eff}v^{(t)}\sum_{i=1}^{2b}(\beta_{i,m}^{(t)})^2c_i}{1+\lambda_\mathrm{eff}v^{(t)}\sum_{i=1}^{2b}(\beta_{i,m}^{(t)})^2}.
\end{equation}
Therefore,
\begin{equation}
    \eta\left({\bm{r}}^{(t)}, v^{(t)}, \bm{b}, \bm{b_v}\right) = \frac{\bm{r}^{(t)}+\lambda_\mathrm{eff}v^{(t)}\bm{b}}{\bm{1}+\lambda_\mathrm{eff}v^{(t)}\bm{b_v}},
    \label{eqn:amp_denoiser}
\end{equation}
where
\begin{equation}
    \bm{b} \triangleq \sum_{i=1}^{2^b} c_i [(\beta_{i,1}^{(t)})^2, \cdots, (\beta_{i,M}^{(t)})^2 ],
    \label{eqn:b_vector}
\end{equation}
\begin{equation}
    \bm{b_v} \triangleq \sum_{i=1}^{2^b}[(\beta_{i,1}^{(t)})^2, \cdots,(\beta_{i,M}^{(t)})^2 ].
    \label{eqn:b_v}
\end{equation}

It is critical, however, to highlight a stark distinction between standard \ac{AMP} theory and our proposed \ac{RAMP} algorithm. 
As mentioned before, the Onsager term in \eqref{eqn:residual_upd} is designed to perfectly cancel the correlation between the current estimate and $\bm{r}^{(t)}$, which ensures that the effective noise is \ac{i.i.d.} Gaussian at each iteration.

However, since the weights $\beta^{(t)}_{i,m}$ are recomputed from $\hat{\bm{s}}^{(t-1)}$ at every iteration, the denoiser has a dependence upon the previous estimate. The cross-iteration dependence causes a residual correlation to accumulate, which the Onsager correction is not able to cancel.
The resulting non-Gaussian effective noise, as shown to be the case in Fig. \ref{fig:QQ_plot}, causes the empirical \ac{MSE} to deviate from the \ac{SE}. This is further confirmed by a controlled experiment in which $\beta^{(t)}_{i,m}$ are computed from the ground-truth $\bm{s}$; under this condition, the \ac{SE} tracks the empirical \ac{MSE} exactly.

Although rectifications are possible through \ac{OAMP} 
\cite{Ma2017,Liu2023} or \ac{VAMP} \cite{Rangan2018, 
Zou2022}, these require \ac{LMMSE} steps involving matrix inversions, re-introducing the $\mathcal{O}(M^3)$ bottleneck. Thus, the memoryless Onsager approximation is a deliberate design choice.

\begin{algorithm}[H]
\caption{RAMP}\label{alg:alg1}
\begin{algorithmic}[1]
\STATE {\textbf{External Input:}}
\STATE \text{Received signal $\bm{y}$, channel matrix $\bm{H}$ \& noise power $\sigma_{\bm{n}}^2$}

\STATE {\textbf{Internal Parameters:}}
\STATE \text{Maximum number of iteration} ${t}_{max}$
\STATE \text{Convergence threshold $\epsilon \ll 1$}, \text{$\alpha \ll 1$}

\STATE \textbf{Initialization:}
\STATE \text{Set initial solution to $\hat{\bm{s}}^{(0)}=\bm{0}_{M\times1}$ and $\bm{z}^{(0)}=\bm{0}_{N\times1}$}
\STATE \textbf{repeat}
\STATE \hspace{0.5cm}\text{Increase iteration counter} $t = t +1$
\STATE \hspace{0.5cm}\text{Update $v^{(t)}$ as in equation \eqref{eqn:variance_upd}}
\STATE \hspace{0.5cm}\text{Update $\bm{r}^{(t)}$ as in equation \eqref{eqn:effective_obs}}
\STATE \hspace{0.5cm}\text{Update} $\beta_{i,m}^{(t)}\forall i,m$ \text{as in equation \eqref{eqn:beta}}
\STATE \hspace{0.5cm}\text{Obtain $\hat{\bm{s}}^{(t+1)}$ as in equation \eqref{eqn:amp_denoiser}}
\STATE \hspace{0.5cm}\text{Update $\bm{z}^{(t+1)}$ as in equation \eqref{eqn:residual_upd}}
\STATE \hspace{0.5cm}\text{Calculate $\varepsilon_k = \|\hat{\bm{s}}^{(t+1)}-\hat{\bm{s}}^{(t)}\|_2$}
\STATE \textbf{until } $t>t_{max}$ or $\varepsilon_t<\varepsilon$
\end{algorithmic}
\end{algorithm}

Despite the deviation, the algorithm maintains robust convergence due to the intrinsic adaptive damping mechanism of the fractional programming formulation. 
When the estimate $\hat{\bm{s}}^{(t)}$ is far from the constellation $\mathcal{S}$, the weights $\beta_{i,m}$ are small.
This suppresses the regularization term in \eqref{eqn:amp_denoiser}, preventing aggressive corrections. 
Similarly, as the estimate settles near a discrete point, $\beta_{i,m}$ increases, effectively locking into the correct symbol.

With the denoiser introduced in \eqref{eqn:amp_denoiser}, we summarize the overall \ac{RAMP} scheme in the form of a pseudocode in Algorithm 1.

\subsection{Robust RAMP Detector}
While the \ac{RAMP} detector derived in the previous section successfully solves \eqref{eqn:IDLS_Base_Case}, as mentioned before, this framework can be readily extended to solve the noise-resilient robust formulation \eqref{eqn:IDLS_Robust_Case}.
The problem is formulated by introducing an $\ell_2$-norm penalty on the signal $\bm{s}$ itself, analogous to the \ac{LMMSE} estimator.

Re-interpreting \eqref{eqn:IDLS_Robust_Case} within our \ac{MAP} framework again requires normalizing the entire objective by the noise variance $\sigma_n^2$ to achieve probabilistically consistent scaling
\begin{equation}
    \min_{s \in \mathbb{C}^{M}} 
    \underbrace{ \frac{1}{\sigma_n^2} \|y - Hs\|_2^2 }_{\propto -\log P(y|s)} + 
    \underbrace{ \|s\|_2^2 + \lambda_\mathrm{eff} \sum_{i=1}^{2^b}\sum_{m=1}^{M} \beta_{i,m}^2 |s_m - c_i|^2 }_{\propto -\log P(s)},
    \label{eqn:normalized_robust_cost_func}
\end{equation}
where $\lambda_\mathrm{eff} \triangleq \lambda/\sigma_n^2$. 

Comparing \eqref{eqn:normalized_robust_cost_func} to \eqref{eqn:normalized_cost_func}, the only change is the addition of the $\|s\|_2^2$ term to the negative log-prior. We can therefore similarly derive the corresponding \ac{AMP} denoiser for this enhanced prior model as
\begin{equation}
    \eta\left(\bm{r}^{(t)}, v^{(t)}, \bm{b}, \bm{b_v}\right) = \frac{\bm{r}^{(t)}+\lambda_\mathrm{eff}v^{(t)}\bm{b}}{\bm{1}+ v^{(t)} +\lambda_\mathrm{eff}v^{(t)}\bm{b_v}},
    \label{eqn:robust_amp_denoiser}
\end{equation}
where $b$ and $b_v$ are the same as defined in \eqref{eqn:b_vector} and \eqref{eqn:b_v}. 

Since the overall \ac{AMP} framework remains identical, the robust \ac{RAMP} detector is implemented by simply replacing the base denoiser \eqref{eqn:amp_denoiser} used in \textit{line 13} of Algorithm 1 with the robust denoiser \eqref{eqn:robust_amp_denoiser}.

\section{Performance Analysis}
We evaluate the proposed \ac{RAMP} and robust \ac{RAMP} algorithms against the high-complexity, \ac{SotA} \ac{IDLS} baselines under an uncorrelated Rayleigh fading channel model.
The elements of the channel matrix $H$ are modeled as \ac{i.i.d.} circularly symmetric complex Gaussian random variables, $h_{m,n} \sim \mathcal{CN}(0,1)$.

\subsection{Numerical Results}
The transmitted symbols are drawn from a Gray-coded \ac{QPSK} constellation ($b=2$ bits per symbol).
The $\ell_0$-norm approximation tightness parameter for the fractional programming weights is fixed at $\alpha = 0.1$. 
The corresponding \ac{AWGN} noise variance for a given $E_b/N_0$ is calculated as $\sigma_n^2 = M / (b \cdot 10^{\frac{E_b/N_0}{10}})$.

\begin{figure}[H]
    \centering
    \includegraphics[width=\columnwidth]{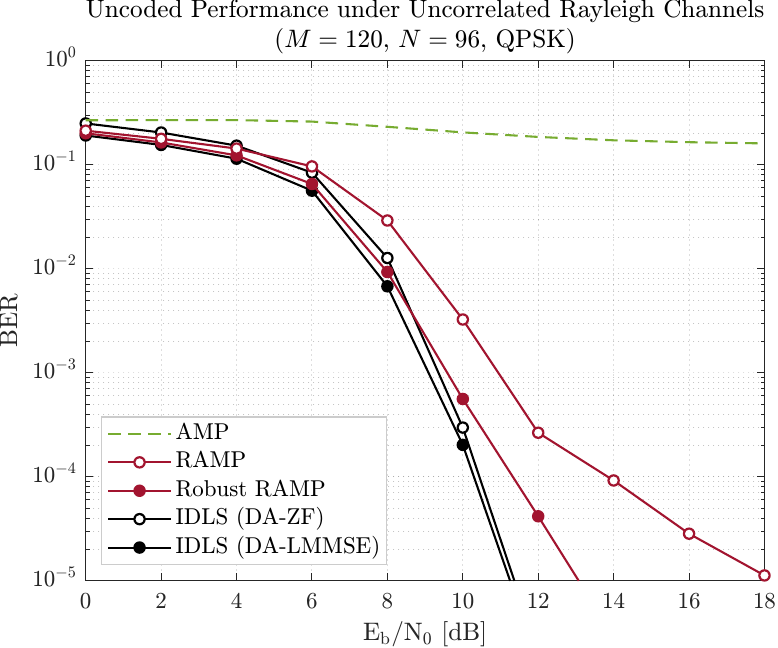}
    \caption{BER vs. SNR performance in overloaded conditions.}
    \label{fig:BER_OverLoaded}
%\vspace{-2ex}
%
    \includegraphics[width=\columnwidth]{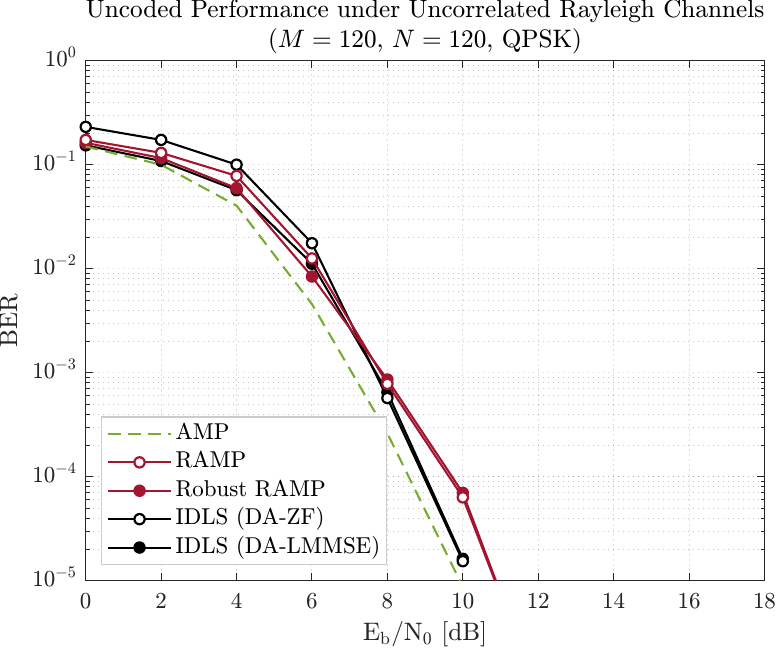}
    \caption{BER vs. SNR performance in fully-loaded conditions.}
    \label{fig:BER_fullyLoaded}
%    \vspace{-2ex}
%
    \centering
    \includegraphics[width=\columnwidth]{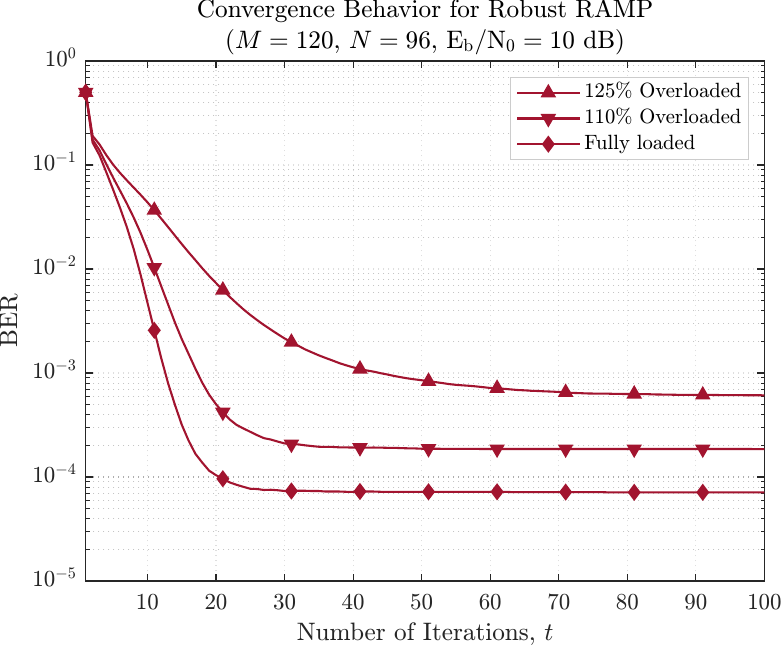}
    \caption{Convergence behavior of the robust RAMP detector.}
    \label{fig:Convergence}
    \vspace{-2ex}
\end{figure}

We investigate both a fully loaded system ($M = 120, N = 120$) and a moderately overloaded system with a 25\% overload ($M = 120, N = 96$).

Figures \ref{fig:BER_OverLoaded} and \ref{fig:BER_fullyLoaded} present the uncoded BER performance versus $E_b/N_0$ for the overloaded and fully loaded scenarios, respectively. 
The \ac{DA-ZF} and \ac{DA-LMMSE} curves represent the exact, closed-form \ac{IDLS} solutions. 
As observed in both scenarios, the proposed \ac{RAMP} and robust \ac{RAMP} algorithms track their respective exact counterparts with good accuracy, exhibiting very little performance degradation. 

Furthermore, as anticipated from the $l_2$-norm penalty, the robust \ac{RAMP} formulation provides superior noise resilience, consistently outperforming the base \ac{RAMP} detector across the simulated SNR range.
Most crucially, Fig. \ref{fig:BER_OverLoaded} highlights the catastrophic failure of standard AMP in the 125\% overloaded regime, where it suffers from an irreducible error floor.
In stark contrast, both proposed RAMP variants successfully exploit the discrete prior to resolve the overloaded scenario, achieving excellent performance.
These results empirically demonstrate that the proposed message-passing framework successfully captures the discrete-aware performance of the $\mathcal{O}(M^3)$ complex \ac{IDLS} solution at the highly efficient complexity of $\mathcal{O}(NM)$. 

As discussed in Section \ref{section:RAMP}, the cross-iteration dependent nature of our derived denoisers means the distribution of effective noise deviates from the purely Gaussian assumption required by standard \ac{AMP} \ac{SE}.
To demonstrate that the proposed algorithm ensures convergence despite this deviation, Fig. \ref{fig:Convergence} illustrates the BER evolution as a function of the iteration number $t$ at a fixed $E_b/N_0 = 10$ dB.
Our proposed \ac{RAMP} algorithms exhibit rapid convergence, reaching convergence within approximately 20-30 iterations.

While only \ac{QPSK} simulations are presented here, the framework is applicable to any finite discrete alphabet, including higher-order \ac{QAM} constellations.
 
\subsection{Complexity Analysis}
A significant contribution of this work is the reduction of computational complexity, enabling the use of the high-performance \ac{IDLS} framework in large-scale systems.
The original \ac{IDLS} framework requires a direct matrix inversion at each iteration, resulting in a prohibitive cubic complexity of $\mathcal{O}(M^3)$. 
By decoupling this inversion, the per-iteration complexity of both proposed \ac{RAMP} variants is dominated solely by the matrix-vector multiplications $H^H z^{(t-1)}$ and $H \hat{s}^{(t-1)}$ in \eqref{eqn:effective_obs} and \eqref{eqn:residual_upd} respectively, reducing the complexity to $\mathcal{O}(NM)$.

% \begin{table}[H]
% \caption{Table of Complexity Order}
% \normalsize
% \centering
% \begin{tabular}{|c|c|c|c|c|}
% \hline
% Method & \makecell{Robust \\ \& RAMP} & \makecell{IDLS}\\
% \hline
% \makecell{Complexity \\ Order} &  $\mathcal{O}(N M)$ & $\mathcal{O}(M^{3})$\\
% \hline
% \end{tabular}
% \end{table}

\section{Conclusion}
In this paper, we addressed the computational bottleneck of discrete-aware symbol detection in overloaded \ac{MIMO} systems. 
While the \ac{SotA} \ac{IDLS} framework achieves near-optimal performance by relaxing the discrete constraints via fractional programming, its iterative matrix inversions impose a prohibitive $\mathcal{O}(M^3)$ complexity.
To overcome this, we proposed the \ac{RAMP} and robust \ac{RAMP} algorithms. 
By deriving an adaptive scalar denoiser that enforces discrete constellation constraints, our framework successfully avoids the high-dimensional matrix inversion entirely. 
Consequently, the proposed methods closely match the exceptional performance of the exact \ac{IDLS} formulations, while drastically reducing the computational complexity to a highly scalable $\mathcal{O}(NM)$.


\begin{thebibliography}{99}

\bibitem{yang2015mimo}
S. Yang and L. Hanzo, 
``Fifty Years of MIMO Detection: The Road to Large-Scale MIMOs,'' 
\textit{IEEE Communications Surveys \& Tutorials}, 
vol. 17, no. 4, pp. 1941--1988, 2015, 
doi: 10.1109/COMST.2015.2475242.

\bibitem{FoucartRauhut2013}
S. Foucart et. al,
\textit{A Mathematical Introduction to Compressive Sensing},
ser. Applied and Numerical Harmonic Analysis. Birkh\"{a}user, 2013.

\bibitem{Das2013}
A.~K. Das and S.~Vishwanath,
``On finite alphabet compressive sensing,''
in \textit{Proc. IEEE Int. Conf. Acoustics, Speech and Signal Processing (ICASSP)},
Vancouver, BC, Canada, May 2013, pp. 5890--5894.

\bibitem{Hayakawa2017}
R. Hayakawa and K. Hayashi,
``Convex Optimization-Based Signal Detection for Massive Overloaded
MIMO Systems,''
\textit{IEEE Trans. Wireless Commun.},
vol. 16, no. 11, pp. 7080--7091, 2017.

\bibitem{Hayakawa2018}
-----,
``Reconstruction of Complex Discrete-Valued Vector via Convex
Optimization with Sparse Regularizers,''
\textit{IEEE Access}, 2018.

\bibitem{Donoho2009}
D. L. Donoho, A. Maleki, and A. Montanari,
``Message-Passing Algorithms for Compressed Sensing,''
\textit{Proc. Nat. Acad. Sci.},
vol. 106, no. 45, pp. 18914--18919, Nov. 2009.

\bibitem{Bayati2011}
M. Bayati and A. Montanari,
``The Dynamics of Message Passing on Dense Graphs,
with Applications to Compressed Sensing,''
\textit{IEEE Trans. Inf. Theory},
vol. 57, no. 2, pp. 764--785, Feb. 2011.

\bibitem{Montanari2011}
A. Montanari,
“Graphical Models Concepts in Compressed Sensing,”
in Compressed Sensing: Theory and Applications,
Y. C. Eldar and G. Kutyniok, Eds.
Cambridge, U.K.: Cambridge Univ. Press, 2012, ch. 9, pp. 394–438,
doi: 10.1017/CBO9780511794308.010.

\bibitem{iimori2020robust}
H. Iimori, G. T. F. de Abreu, T. Hara, K. Ishibashi, R.-A. Stoica,
D. G. Gonzalez, and O. Gonsa,
“Robust Symbol Detection in Large-Scale Overloaded NOMA Systems,”
IEEE Open Journal of the Communications Society, vol. 2, pp. 512–533, 2021,
Art. no. 9373674, doi: 10.1109/OJCOMS.2021.3064983.

\bibitem{Shen2018}
K. Shen and W. Yu,
``Fractional Programming for Communication Systems---Part I:
Power Control and Beamforming,''
\textit{IEEE Trans. Signal Process.},
vol. 66, no. 10, pp. 2616--2630, May 2018.

\bibitem{Boukari1995}
D. Boukari and A. V. Fiacco,
``Survey of Penalty, Exact-Penalty and Multiplier Methods from 1968 to 1993,''
\textit{Optimization}, vol. 32, no. 4, pp. 301--334, 1995.
doi:10.1080/02331939508844023.

\bibitem{Pedregosa2016}
F. Pedregosa,
``Hyperparameter Optimization with Approximate Gradient,''
in \textit{Proceedings of the 33rd International Conference on Machine Learning (ICML)},
vol. 48, Proceedings of Machine Learning Research, pp. 737--746, 2016.

\bibitem{Ehrhardt2024}
M. J. Ehrhardt, S. Gazzola, and S. J. Scott, ``On Optimal Regularization Parameters via Bilevel Learning,''
in \textit{Radon Series on Computational and Applied Mathematics},
De Gruyter, 2024.

\bibitem{Bischl2023}
B. Bischl, M. Binder, M. Lang, B. Pfahringer, J. Richter, and D. Surmann, ``Hyperparameter Optimization: Foundations, Algorithms, Best Practices, and Open Challenges,''
\textit{WIREs Data Mining and Knowledge Discovery}, vol. 13, no. 2, e1484, 2023.
doi:10.1002/widm.1484.

\bibitem{Ma2017}
J. Ma and L. Ping,
“Orthogonal AMP,”
IEEE Access, vol. 5, pp. 2020–2033, 2017,
doi: 10.1109/ACCESS.2017.2673542.

\bibitem{Liu2023}
L.~Liu, Y.~Cheng, S.~Liang, J.~H. Manton, and L.~Ping,
``On {OAMP}: Impact of the orthogonal principle,''
\textit{IEEE Trans. Commun.},
vol.~71, no.~5, pp.~2992--3007, May 2023.

\bibitem{Rangan2018}
S.~Rangan, P.~Schniter, and A.~K. Fletcher,
``Vector approximate message passing,''
\textit{IEEE Trans. Inf. Theory},
vol.~65, no.~9, pp.~5339--5351, Sep. 2019.

\bibitem{Zou2022}
Q. Zou and H. Yang,
``A Concise Tutorial on Approximate Message Passing,''
2022. [Online]. Available: https://arxiv.org/abs/2201.07487

\end{thebibliography}
\end{document}